\documentclass[twocolumn,showpacs,floatfix]{revtex4}

\usepackage{graphicx}

\begin{document}

\title{Spatial, spectral and temporal coherence of ultra-intense twin
beams}

\author{Jan Pe\v{r}ina Jr.}
\email{jan.perina.jr@upol.cz}
\affiliation{RCPTM, Joint Laboratory
of Optics of Palack\'{y} University and Institute of Physics of
Academy of Sciences of the Czech Republic, Faculty of Science,
Palack\'{y} University, 17. listopadu 12, 77146 Olomouc, Czech
Republic}

\begin{abstract}
Using the model of parametric interaction based on the
spatio-spectral Schmidt modes and generalized parametric
approximation, we analyze coherence and mode structure of
ultra-intense twin beams generated in the regime with pump
depletion. We show that the increase of spatial and spectral
coherence with the increasing pump power observed for moderate
powers is replaced by the decrease for the pump powers at which
pump depletion occurs. This behavior of coherence is opposed to
that exhibited by the number of spatio-spectral modes effectively
constituting the twin beam. The conditions for maximal coherence
are analyzed considering pump-beam parameters (spectral width,
transverse radius). The existence of additional coherence maxima
occurring at even higher pump powers is predicted and explained by
the oscillatory evolution of the modes' populations.
\end{abstract}



\maketitle

\section{Introduction}

Parametric down-conversion (PDC) has been used the most frequently
in its 'weak' spontaneous regime where it provides entangled
photon pairs with genuine quantum properties. On the other hand,
it provides the so-called twin beams containing a large number of
photon pairs when intensively pumped. Such twin beams exhibit
ideally perfect correlations in the photon numbers (or
intensities) of the signal and idler fields that constitute the
twin beam
\cite{Jedrkiewicz2004,Bondani2007,Blanchet2008,Brida2009a}. These
correlations in photon numbers occur in the spectrum as well as in
the fields' transverse planes as a consequence of energy
conservation and phase-matching conditions, respectively. However,
as pulsed pumping is needed to generate intense twin beams and the
nonlinear interaction is restricted to a final volume of the
nonlinear material, neither spectral nor spatial correlations
inside the twin beam are ideal. For this reason, we introduce
spectral and spatial intensity correlation functions to
characterize properties of the real twin beams.

The spectral and spatial intensity correlation functions of twin
beams have been experimentally analyzed in three regions differing
by intensity. In each of them, different behavior has been
observed. It has been shown that the coherence of individual
photon pairs generated in the spontaneous regime is preserved even
for relatively intense twin beams \cite{Machulka2014,Allevi2014}.
In this first region, the pump field is sufficiently intense
relative to the overall strength of the nonlinear interaction
(including the interaction length) so that no pump depletion
occurs. The twin beam is thus composed of a large number of weakly
populated signal and idler modes.

An increase in the spatial and spectral coherence of a twin beam
is characteristic for the second region
\cite{Christ2011,Spasibko2012,Christ2013,Perez2014,Brecht2014,Sharapovova2015}.
The theory and its comparison with the experimentally determined
numbers of twin-beam modes have revealed that greater mean photon
numbers in individual signal and idler Schmidt modes occur in this
region \cite{PerinaJr2015a}. Moreover, the mean photon numbers of
the most intense signal and idler modes are such that the pump
depletion has to occur during the interaction. The increase of
coherence observed in this region has been explained by the
dominance of the Schmidt modes with the greatest Schmidt
coefficients over the other Schmidt modes. The reason is that the
number of highly-populated modes is smaller and so the twin beam
exhibits better coherence properties. Decrease in the number of
modes accompanying the increase of coherence has been
experimentally confirmed in \cite{Allevi2014a}.

Finally, the third region reached with ultra-intense pump beams
has been investigated experimentally only recently
\cite{Allevi2014a,Allevi2015,Allevi2015a}. Here, the coherence of
the twin beam decreases whereas the number of modes increases. As
shown in this paper, this behavior can be explained, similarly as
in the second region, by considering the evolution of mean photon
numbers of the individual signal and idler Schmidt modes. However,
in this region the signal and idler Schmidt modes with the
greatest Schmidt coefficients already loose their energy during
the propagation, in favor of the pump modes that originally
provided their energy at the beginning of the nonlinear
interaction.

The three intensity regions are easily identified when the
twin-beam intensity is considered as a function of pump power. An
initial exponential increase of the twin-beam intensity occurs in
the first region. The area where the exponential increase is
gradually replaced by the linear one forms the second region.
Finally, the pump powers at which a nearly linear increase of
twin-beam intensity is found belong to the third region. The
values of pump powers at which such behavior is observed depend
strongly on the properties of nonlinear medium. It holds in
general that the smaller is the number of spatio-spectral modes,
the smaller are the pump powers. Also, the stronger is the
nonlinear interaction, the smaller are the pump powers. From this
point of view, the pump powers lower by several orders in
magnitude are expected in nonlinear photonic structures (e.g.,
waveguides) compared to bulk crystals analyzed in this paper.

Here, we develop the theory explaining the behavior of intense
twin beams in the third intensity region, i.e. when the pump
depletion is substantial for twin-beam properties. We introduce
\emph{the generalized parametric approximation} in which the pump
beam is depleted during its propagation in accordance with the
classical solution of the nonlinear interaction. We decompose the
twin beam together with the pump beam into multiple triplets of
individual spatio-spectral modes defined in the signal, idler and
pump beams. We adopt the signal and idler spatio-spectral Schmidt
modes defined for weak twin beams
\cite{Law2000,Law2004,Fedorov2014,PerinaJr2015} following the
approach elaborated in \cite{PerinaJr2015a}. The presented theory
represents a generalization of the theory given in
\cite{PerinaJr2015a} that assumes un-depleted pump beams. As such,
it also covers the first and the second intensity regions.

The behavior of twin beams in the region with pump depletion has
been successfully described replacing the initial vacuum quantum
state of a twin beam by a classical statistical ensemble and then
finding the nonlinear evolution numerically
\cite{Allevi2014a,Brambilla2004}. The loss of coherence for high
pump powers as well as spectral and spatial deformations of the
pump beam have been revealed in this approach. On the other hand,
the relation between the coherence and the twin-beam internal
structure cannot be analyzed in this approach. For this reason, we
generalize the model of Ref.~\cite{PerinaJr2015a} based on the
Schmidt modes to the region with pump depletion. This allows us,
among others, to predict additional coherence maxima found at high
pump powers.

We note that a theory of intense twin beams based upon the
quasi-monochromatic and quasi-plane-wave pump-field approximations
has been developed for the pump powers belonging to the second
region \cite{Gatti2003,Brambilla2010,Caspani2010,Dayan2007}.
Similarly as the presented theory, this theory has been based on
the solution of the linear Heisenberg equations. The suggested
generalized parametric approximation could be applied also here
with the potential to generalize the existing theory into the
region with pump depletion.

The developed theory is suitable not only for parametric
interactions \cite{Boyd2003}. It can be applied also to nonlinear
resonant interactions involving four-wave mixing in cold atomic
ensembles \cite{Kolchin2006,Glorieux2010,Boyer2008} in which high
effective nonlinear coupling constants occur.

The paper is organized as follows. In Sec.~II, the theory of
intense twin beams in the pump-depleted regime is developed and
quantities characterizing twin beams are defined. Spatial,
spectral and temporal coherence of twin beams is discussed in
Sec.~III. Multiple coherence maxima observed with the increasing
pump power and their relation to the twin-beam structure are
analyzed in Sec.~IV. A model with an extended interaction length
is briefly discussed in Sec.~V. Conclusions are drawn in Sec.~VI.

\section{The model of ultra-intense twin beams involving pump depletion}

An intense twin beam is generated in the process of PDC in which
the pump field interacts nonlinearly with the signal and idler
fields. The signal and idler fields are assumed initially in the
vacuum states. They take energy from the pump field during the
interaction. The nonlinear interaction mediated by the material
with $ \chi^{(2)} $ nonlinearity is described by the following
nonlinear interaction momentum operator $ \hat{G}_{\rm int} $
\cite{Boyd2003,Perina1991,PerinaJr2000}:
\begin{eqnarray}   
 \hat{G}_{\rm int}(z) &=& 2 \epsilon_0 \int dxdy \int dt \nonumber \\
 & & \hspace{-1cm} \left[
  \chi^{(2)} \hat{E}^{(+)}_{\rm p}({\bf r},t) \hat{E}^{(-)}_{\rm s}({\bf r},t)
  \hat{E}^{(-)}_{\rm i}({\bf r},t) + {\rm h.c.} \right];
\label{1}
\end{eqnarray}
$ {\bf r} = (x,y,z) $. In Eq.~(\ref{1}), $ \hat{E}^{(+)}_{\rm p} $
denotes the positive-frequency pump electric-field operator
amplitude whereas $ \hat{E}^{(-)}_{\rm s} $ [$ \hat{E}^{(-)}_{\rm
i} $] stands for the negative-frequency part of the signal [idler]
electric-field operator amplitude. The vacuum permittivity is
denoted as $ \epsilon_0 $ and $ {\rm h.c.} $ replaces the
Hermitian conjugated term. The electric-field amplitudes $
\hat{E}_a $, $ a={\rm p,s,i} $, are assumed to be decomposed into
the basis of monochromatic plane waves that have their photon
annihilation and creation operators.

To reveal a spatio-spectral structure of the fields suitable for
describing the nonlinear interaction we first consider the
generation of individual photon pairs in the weak nonlinear
interaction. In this case, the pump field is treated classically
and an emitted photon pair is described by a quantum state $
|\psi\rangle_{\rm si} $ obtained by the first-order perturbation
solution of the corresponding Schr\"{o}dinger equation. This
solution allows us to find suitable spatio-spectral Schmidt dual
modes (identified by triple indices $ mlq $) and then to associate
with them the signal- and idler-field creation operators $
\hat{a}_{{\rm s},mlq}^\dagger $ and $ \hat{a}_{{\rm
i},mlq}^\dagger $, respectively. Assuming for simplicity the
factorization of modes into their spatial (indices $ ml $) and
spectral ($ q $) parts, the state $ |\psi\rangle_{\rm si} $ is
written as (for more details, see \cite{PerinaJr2015a}):
\begin{eqnarray}  
 |\psi\rangle_{\rm si} = t^\perp f^\parallel \sum_{m,l,q} \lambda_{ml}^\perp
  \lambda_q^\parallel \hat{a}_{{\rm s},mlq}^{\dagger}
   \hat{a}_{{\rm i},mlq}^{\dagger} |{\rm vac}\rangle ;
\label{2}
\end{eqnarray}
$ |{\rm vac}\rangle $ stands for the signal- and idler-field
vacuum state. The coefficients $ \lambda_{ml}^\perp $ ($
\lambda_q^\parallel $) give the probability amplitudes of having
an $ ml $-th ($ q $-th) spatial (spectral) mode in the generated
state. Symbol $ t^\perp $ ($ f^\parallel $) denotes an appropriate
spatial (spectral) normalization constant. The Schmidt-mode
creation operators $ \hat{a}_{b,mlq}^\dagger $ occurring in
Eq.~(\ref{2}) are given as linear combinations of the
monochromatic (frequency $ \omega_b $) plane-wave (radial
transverse wave-vector coordinates $ k_b^\perp $ and $ \varphi_b
$) mode creation operators $
\hat{a}_{b}^\dagger(k_b^\perp,\varphi_b,\omega_b) $:
\begin{eqnarray}  
 \hat{a}_{b,mlq}^\dagger &=& \int_{0}^{\infty} dk_b^\perp \int_{0}^{2\pi}d\varphi_b
   \int_{0}^{\infty} d\omega_b \, t_{b,ml}(k_b^\perp,\varphi_b)  \nonumber \\
 & & \mbox{} \times  f_{b,q}(\omega_b) \hat{a}_{b}^\dagger(k_b^\perp,\varphi_b,\omega_b),\,
 \hspace{5mm} b={\rm s,i}.
\label{3}
\end{eqnarray}
In Eq.~(\ref{3}), the Schmidt mode functions $ t_{b,ml} $ and $
f_{b,q} $ defined in the beam's transverse wave-vector plane
\cite{Fedorov2008,Fedorov2007} and the frequency domain
\cite{Mikhailova2008}, respectively, have been used.

Modes in the pump field can be assigned to individual pairs of the
signal and idler Schmidt modes obtained at the single photon-pair
level. This results in mutually independent modes' triplets. They
can be conveniently used for rewriting the momentum operator $
\hat{G}_{\rm int} $ in the approximative form:
\begin{eqnarray}     
 \hat{G}_{\rm int}^{\rm av}(z) &=& - i\hbar K
  \sum_{m=-\infty}^{^\infty} \sum_{l,q=0}^{\infty}
   \hat{a}_{{\rm p},mlq}(z) \hat{a}_{{\rm s},mlq}^{\dagger}(z)
   \hat{a}_{{\rm i},mlq}^{\dagger}(z)\nonumber \\
 & & \mbox{}  + {\rm h.c.};
\label{4}
\end{eqnarray}
$ \hbar $ stands for the reduced Planck constant. A common
coupling constant $ K $ introduced in Eq.~(\ref{4}) includes
multiplicative factors $ t^\perp f^\parallel $ quantifying the
strength of nonlinear interaction in the medium of length $ L $
and normalization with respect to photon numbers ($ \xi_{\rm p}
$); $ K = t^\perp f^\parallel/(L\xi_{\rm p}) $. The pump power $
P_{\rm p} $, its repetition rate $ f $ and its central frequency $
\omega_{\rm p}^0 $ determine the overall pump-field amplitude $
\xi_{\rm p} $ in the form $ \xi_{\rm p} = \sqrt{ P_{\rm p}
/(f\hbar\omega_{\rm p}^0)} $. It is assumed that the pump power $
P_{\rm p} $ can be divided into individual pump modes indexed by $
mlq $ linearly proportionally to their squared Schmidt
coefficients $ (\lambda_{ml}^\perp \lambda_q^\parallel)^2 $. This
means that an initial classical (coherent) amplitude $ A_{{\rm
p},mlq}^{\cal N}(0) = \lambda_{ml}^\perp \lambda_q^\parallel
\xi_{\rm p} $ is assigned to an $ (mlq)$-th mode.

The Heisenberg equations derived from the momentum operator $
\hat{G}_{\rm int}^{\rm av} $ in Eq.~(\ref{4}) are written for
individual modes' triplets as follows:
\begin{eqnarray}   
 \frac{ d\hat{a}_{{\rm s},mlq}(z)}{dz} &=&
   K\hat{a}_{{\rm p},mlq}(z) \hat{a}_{{\rm i},mlq}^\dagger(z) , \nonumber \\
 \frac{ d\hat{a}_{{\rm i},mlq}(z)}{dz} &=&
   K\hat{a}_{{\rm p},mlq}(z) \hat{a}_{{\rm s},mlq}^\dagger(z), \nonumber \\
 \frac{ d\hat{a}_{{\rm p},mlq}(z)}{dz} &=&
   - K\hat{a}_{{\rm s},mlq}(z) \hat{a}_{{\rm i},mlq}(z).
\label{5}
\end{eqnarray}
The operator equations (\ref{5}) are nonlinear. They can be solved
exactly only for the interacting fields with small photon numbers
invoking numerical approach. Here, we find an approximative
solution using the fact that the pump field remains strong during
the interaction. We treat it classically and express its evolution
along the nonlinear medium using the solution of classical
nonlinear equations. We further pay attention to one typical
modes' triplet and omit the indices $ mlq $ for simplicity. We
transform the Heisenberg operator equations (\ref{5}) into their
classical analog written for fields' amplitudes determined for the
symmetric ordering of fields' operators:
\begin{eqnarray}  
 \frac{dA_{\rm s}(z)}{dz} &=& K A_{\rm p}(z) A_{\rm i}(z) ,
  \nonumber \\
 \frac{dA_{\rm i}(z)}{dz} &=& K A_{\rm p}(z) A_{\rm s}(z) ,
  \nonumber \\
 \frac{dA_{\rm p}(z)}{dz} &=& - K A_{\rm s}(z)A_{\rm i}(z).
\label{6}
\end{eqnarray}
In the symmetric ordering, $ A_{\rm p}(0) = \sqrt{(A_{\rm
p}^{{\cal N}})^2(0) + 1/2} $ and $ A_{\rm s}(0) = A_{\rm i}(0) =
1/\sqrt{2} $ for the initial vacuum signal- and idler-field
amplitudes.

As the signal and idler fields occur symmetrically in PDC and they
both begin the interaction in the vacuum state, their classical
amplitudes are equal, i.e. $ A_{\rm s}(z) = A_{\rm i}(z) $. This
together with the integral of motion $ A_{\rm p}^2(z) + A_{\rm
s}^2(z) = A_{\rm p}^2(0) + A_{\rm s}^2(0) \equiv A_{\rm ps}^2 $
and assumption of real amplitudes $ A_{\rm p} $, $ A_{\rm s} $ and
$ A_{\rm i} $ and real coupling constant $ K $ allows to solve
Eqs.~(\ref{6}) analytically. In detail, the third equation in
(\ref{6}) is transformed into a differential equation for $ A_{\rm
p}(z) $ that can be solved by direct integration. The integral of
motion then provides the signal- and idler-field amplitudes $
A_{\rm s}(z) $ and $ A_{\rm i}(z) $:
\begin{eqnarray}   
 A_{\rm p}(z) &=& A_{\rm ps} \frac{A_{\rm p}{\rm cosh}(KA_{\rm ps}z)- A_{\rm ps}{\rm
  sinh}(KA_{\rm ps}z) }{ A_{\rm ps}{\rm
  cosh}(KA_{\rm ps}z) - A_{\rm p}{\rm sinh}(KA_{\rm ps}z)} ,
\label{7}   \\
 A_{\rm s}(z) &=&  \frac{A_{\rm s} A_{\rm ps}}{ A_{\rm ps}{\rm cosh}(KA_{\rm ps}z) -A_{\rm p}{\rm
  sinh}(KA_{\rm ps}z)};
\label{8}
\end{eqnarray}
$ A_{\rm p} \equiv A_{\rm p}(0) $ and $ A_{\rm s} \equiv A_{\rm
s}(0) $. The solution written in Eqs.~(\ref{7}) and (\ref{8}) has
been obtained for the nonzero pump field $ A_{\rm p}(z) $. The
pump field is completely depleted at $ z=z_0 $ for which $ A_{\rm
p}(z_0) = A_{\rm s} = 1/\sqrt{2} $:
\begin{equation}   
 z_0 = \frac{1}{ 2KA_{\rm ps} } \ln \left[ 1 + \frac{2A_{\rm ps}}{A_{\rm ps}+A_{\rm s}}
  \frac{A_{\rm p} - A_{\rm s}}{A_{\rm ps}-A_{\rm p}} \right] .
\label{9}
\end{equation}
At this point, the phases of the interacting fields change such
that the pump field begins to take its energy back from the signal
and idler fields. At the point $ z = 2z_0 $ all energy is back in
the pump field and the evolution repeats. In the interval $ z_0
\le z \le 2z_0 $, the fields' evolution is again described by
Eqs.~(\ref{7}) and (\ref{8}) with the variable $ z $ substituted
by $ 2z_0 -z $.

After finding the classical solution we return back to the
operator equations (\ref{5}). We assume the pump field in the
classical form described in Eq.~(\ref{7}). This generalizes the
usual parametric approximation in which the pump field is treated
as a constant. We call this approach as \emph{the generalized
parametric approximation}. The Heisenberg equations for the
signal- and idler-field annihilation operators then form a linear
system of operator equations:
\begin{eqnarray}  
 \frac{d \hat{a}_{\rm s}(z)}{dz} &=& K A_{\rm p}(z) \hat{a}_{\rm i}^\dagger(z),
  \nonumber \\
 \frac{d \hat{a}_{\rm i}(z)}{dz} &=& K A_{\rm p}(z) \hat{a}_{\rm s}^\dagger(z).
\label{10}
\end{eqnarray}
The solution of Eqs.~(\ref{10}) can be found for an arbitrary
pump-field profile $ A_{\rm p} (z) $ in the form generalizing that
found in the parametric approximation:
\begin{eqnarray}  
 \hat{a}_{\rm s}(z) &=& U(z) \hat{a}_{\rm s}(0) + V(z) \hat{a}_{\rm i}^\dagger(0) , \nonumber \\
 \hat{a}_{\rm i}(z) &=& U(z) \hat{a}_{\rm i}(0) + V(z) \hat{a}_{\rm s}^\dagger(0)
\label{11}
\end{eqnarray}
and
\begin{equation}  
 U(z) = {\rm cosh}[\varphi(z)], \hspace{2mm} V(z) = {\rm
 sinh}[\varphi(z)];
\label{12}
\end{equation}
$ \varphi(z) = \int_{0}^{z} dz' K A_{\rm p}(z') $. We have for the
pump-field amplitude $ A_{\rm p}(z) $ given in Eq.~(\ref{7}):
\begin{equation}   
 \varphi(z) = K A_{\rm ps}z - \ln\left[ \frac{A_{\rm ps}+A_{\rm p}}{2A_{\rm ps}} +
  \frac{A_{\rm ps}-A_{\rm p}}{2A_{\rm ps}} \exp(2KA_{\rm ps}z) \right].
\label{13}
\end{equation}
We note that $ \varphi(z) = K A_{\rm ps}z $ in the usual
parametric approximation. We also note that, similarly as the
parametric approximation, the generalized parametric approximation
does not conserve the energy during the nonlinear interaction. In
fact the energy in the signal (and idler) field given by the
solution (\ref{11}) is lower than that derived classically from
Eq.~(\ref{8}). However, it can be shown that the intensity auto-
and cross-correlation functions as well as numbers of modes are
practically unaffected by this drawback.

The solution (\ref{11}) allows us to determine physical quantities
characterizing twin beams. Here, as an example, we define suitable
quantities in the frequency domain. The definitions of temporal
quantities as well as quantities in the transverse wave-vector
plane are analogous. They can be found in \cite{PerinaJr2015a}.

The signal-field intensity spectrum $ n_{{\rm s},\omega} $
averaged over the transverse modes is defined as follows:
\begin{eqnarray}   
 n_{{\rm s},\omega}(\omega_{\rm s}) &=& \langle \hat{a}_{\rm s}^\dagger(\omega_{\rm s},L)
  \hat{a}_{\rm s}(\omega_{\rm s},L) \rangle_\perp \nonumber \\
 &=& \sum_{ml} \sum_{q} |f_{{\rm s},q}(\omega_{\rm s})|^2 V_{mlq}^2 .
\label{14}
\end{eqnarray}
Symbol $ \langle \rangle_\perp $ stands for quantum mechanical
averaging combined with averaging in the transverse wave-vector
plane. The number $ N_{\rm s} $ of generated signal photons is
easily obtained by the formula
\begin{eqnarray}   
 N_{\rm s} &=& \int_{0}^{\infty} d\omega_{\rm s} \, n_{{\rm s},\omega}(\omega_{\rm s}) =
 \sum_{ml} \sum_{q} V_{mlq}^2 .
\label{15}
\end{eqnarray}

The averaged signal-field spectral intensity correlations are
described by the fourth-order correlation function $ A_{{\rm
s},\omega} $ defined as:
\begin{eqnarray}   
 A_{{\rm s},\omega}(\omega_{\rm s},\omega'_{\rm s}) &=& \langle {\cal N}: \Delta[
  \hat{a}_{\rm s}^\dagger(\omega_{\rm s},L)\hat{a}_{\rm s}(\omega_{\rm s},L)]
  \nonumber \\
 & & \mbox{} \times \Delta[\hat{a}_{\rm s}^\dagger(\omega'_{\rm s},L)
  \hat{a}_{\rm s}(\omega'_{\rm s},L)]:\rangle_{\perp} \nonumber \\
 &=& \sum_{ml} \left|
  \sum_{q} f_{{\rm s},q}^*(\omega_{\rm s}) f_{{\rm s},q}(\omega'_{\rm s})  V_{mlq}^2
  \right|^2 ;
\label{16}
\end{eqnarray}
symbol $ {\cal N}: : $ means the normal ordering of fields'
operators.

Similarly, the spectral intensity cross-correlations between the
signal and idler fields are described using the following
fourth-order correlation function:
\begin{eqnarray}   
 C_{\omega}(\omega_{\rm s},\omega_{\rm i}) &=& \langle {\cal N}: \Delta[
  \hat{a}_{\rm s}^\dagger(\omega_{\rm s},L) \hat{a}_{\rm s}(\omega_{\rm s},L)] \nonumber
  \\
 & & \mbox{} \times  \Delta[ \hat{a}_{\rm i}^\dagger(\omega_{\rm i},L)
  \hat{a}_{\rm i}(\omega_{\rm i},L)]: \rangle_\perp \nonumber \\
 &=& \sum_{ml} \left|
  \sum_{q} f_{{\rm s},q}(\omega_{\rm s}) f_{{\rm i},q}(\omega_{\rm i}) U_{mlq} V_{mlq}
  \right|^2 . \nonumber \\
 & &
\label{17}
\end{eqnarray}

We determine entanglement dimensionality $ K $ of the twin beam
\cite{Gatti2012,Horoshko2012} according to the formula
\begin{eqnarray}   
 K &=& \frac{ \left(\sum_{mlq} \langle \hat{a}_{{\rm s},mlg}(L) \hat{a}_{{\rm i},mlg}(L)
  \rangle^2 \right)^2 }{ \sum_{mlq} \langle \hat{a}_{{\rm s},mlg}(L) \hat{a}_{{\rm i},mlg}(L)
  \rangle^4 } \nonumber \\
 &=& \frac{ \left(\sum_{mlq} U_{mlq}^2V_{mlq}^2 \right)^2 }{
  \sum_{mlq} U_{mlq}^4V_{mlq}^4 }
\label{18}
\end{eqnarray}
that arises after defining suitable photon-pair creation and
annihilation operators \cite{PerinaJr2013}. It quantifies the
number of paired modes found in a twin beam. We note that formula
(\ref{18}) reduces to the usual Schmidt number \cite{Fedorov2014}
for weak twin beams. We also note that several quantifiers of
entanglement dimensionality have been compared in
\cite{Stobinska2012,Chekhova2015,PerinaJr2015a}.

The number of paired modes effectively present in any physical
variable of the twin beam can be estimated by the corresponding
Fedorov ratio $ K^\Delta $ \cite{Fedorov2005}. It is defined as
the ratio of the width $ \Delta n_{\rm s} $ of the signal-field
intensity profile and the width $ \Delta C_{\rm s} $ of the
corresponding intensity cross-correlation function in a given
variable, i.e.
\begin{equation} 
 K^\Delta = \frac{\Delta n_{\rm s}}{\Delta C_{\rm s}} .
\label{19}
\end{equation}

In the following discussion, we consider a pump field with
Gaussian spectrum and Gaussian transverse profile. Its
positive-frequency electric-field amplitude $ E^{(+)}_{\rm
p}(x,y,0,t) $ in front of the crystal is expressed as follows:
\begin{eqnarray}   
 E^{(+)}_{\rm p}(x,y,0,t) &=& \sqrt{\frac{P_{\rm p}}{\epsilon_0 c f} }
 \sqrt{\frac{2}{\pi}}
 \frac{1}{w_{\rm p}} \exp\left[ - \frac{x^2 + y^2}{w_{\rm p}^2} \right]  \nonumber \\
 & &  \hspace{-2.5cm} \mbox{} \times \sqrt{ \sqrt{\frac{2}{\pi} } \frac{1}{\tau_{\rm p}} }
 \exp\left[  - \frac{(1+ia_p) t^2}{\tau_{\rm p}^2} \right] \exp(-i\omega_p^0
 t).
\label{20}
\end{eqnarray}
In Eq.~(\ref{20}), $ w_p $ gives the pump-beam radius, $ \tau_{\rm
p} $ stands for the pump-pulse duration and $ a_{\rm p} $ is the
pump-pulse chirp parameter. Symbol $ c $ denotes the speed of
light in vacuum.

\section{Spatial, spectral and temporal coherence of twin beams}

To demonstrate the behavior of twin beams when the pump power $
P_{\rm p} $ varies, we consider two BBO crystals 4-mm and 8-mm
long both cut for non-collinear type-I process (eoo) for the
spectrally-degenerate interaction pumped by the pulses at
wavelength $ \lambda_{\rm p} = 349 $~nm generated with the
repetition rate $ f = 400 $~s$ {}^{-1} $. These pulses are
experimentally produced by the third harmonics of the Nd:YLF laser
operating at the wavelength 1.047~$\mu $m. Assuming the pump field
at normal incidence, the signal and idler fields at the central
wavelengths $ \lambda_{\rm s}^0 = \lambda_{\rm i}^0 = 698 $~nm ($
\vartheta_{\rm BBO} = 36.3 $~deg) propagate outside the crystal
under the radial emission angles $ \vartheta_{\rm s}^0 =
\vartheta_{\rm i}^0 = 8.45 $~deg. The spectral and transverse
wave-vector Schmidt modes have been determined in
\cite{PerinaJr2015} and applied for the analysis of twin beams
assuming un-depleted pump beams in \cite{PerinaJr2015a}. The
analyzed configuration is symmetric for the signal and idler
fields. That is why we further pay attention to only the
signal-field  properties, together with the joint signal-idler
fields' properties. In this section, we first analyze the
evolution of photon-pair numbers and effective numbers of
twin-beam modes (Subsec. A). Then, we study spectral (Subsec. B),
spatial (Subsec. C) and temporal (Subsec. D) coherence of the twin
beams.

\subsection{Photon-pair numbers, numbers of modes}

\begin{figure}         
 \resizebox{.85\hsize}{!}{\includegraphics{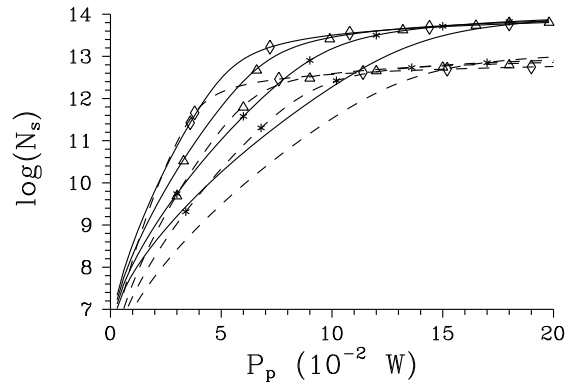}}
  \caption{Mean numbers $ N_{\rm s} $ of signal photons emitted from
   4-mm (solid curves) and 8-mm (dashed curves) long crystals as functions
   of pump power $ P_{\rm p} $ for $ \Delta\lambda_{\rm p} = 1 \times 10^{-9} $~m (plain curves),
   $ \Delta\lambda_{\rm p} = 7 \times 10^{-10} $~m (curves with $ \ast $),
   $ \Delta\lambda_{\rm p} = 5 \times 10^{-10} $~m (curves with $ \triangle $),
   and $ \Delta\lambda_{\rm p} = 3 \times 10^{-10} $~m (curves with $ \diamond $); $ \log $ denotes decimal logarithm;
   $ w_{\rm p} = 1 \times 10^{-3} $~m.}
\label{fig1}
\end{figure}

\begin{figure}         
 \resizebox{.85\hsize}{!}{\includegraphics{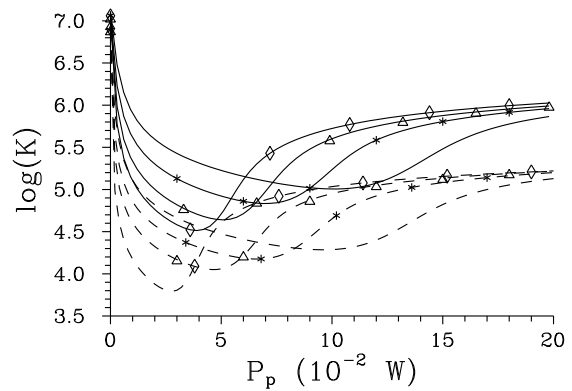}}
  \caption{Numbers $ K $ of twin-beam modes for
   4-mm (solid curves) and 8-mm (dashed curves) long crystals as
   they depend on pump power $ P_{\rm p} $ for $ \Delta\lambda_{\rm p} = 1 \times 10^{-9} $~m (plain curves),
   $ \Delta\lambda_{\rm p} = 7 \times 10^{-10} $~m (curves with $ \ast $),
   $ \Delta\lambda_{\rm p} = 5 \times 10^{-10} $~m (curves with $ \triangle $),
   and $ \Delta\lambda_{\rm p} = 3 \times 10^{-10} $~m (curves with $ \diamond $); $ \log $ denotes decimal logarithm;
   $ w_{\rm p} = 1 \times 10^{-3} $~m.}
\label{fig2}
\end{figure}

The number $ N_{\rm s} $ of emitted photon pairs naturally
increases with the increasing pump power $ P_{\rm p} $. Whereas
the increase is exponential for smaller values of power $ P_{\rm
p} $, this increase is gradually replaced by the linear one for
greater powers $ P_{\rm p} $ (see Fig.~\ref{fig1}). This change
originates in the fact that the mean photon numbers of the
down-converted modes with the greatest Schmidt coefficients $
\lambda $, that have the initial fastest growth and thus form the
initial exponential growth, become saturated and even loose their
energy, due to pump depletion. The initial fast exponential growth
leads to the dominance of the modes with the greatest Schmidt
coefficients $ \lambda $ over the other modes. This reduces the
number $ K $ of twin-beam modes (see Fig.~\ref{fig2}). When this
dominance is lost at greater pump powers $ P_{\rm p} $ due to pump
depletion, the number $ K $ of modes begins to increase for the
powers $ P_{\rm p} $ greater than a certain threshold power $
P_{\rm p,th} $. At this power, the down-converted modes with the
greatest Schmidt coefficients $ \lambda $ begin to send their
energy back to the corresponding pump modes whereas the modes with
smaller Schmidt coefficients $ \lambda $ still take energy from
their pump modes. This assures the continuation of the increase of
the overall energy in the down-converted fields.

For a given crystal, the smaller the number $ K $ of modes
(depending on pump-beam parameters), the faster the increase of
the number $ N_{\rm s} $ of emitted photon pairs with the power $
P_{\rm p} $. This originates from the fact that when dividing the
overall power $ P_{\rm p} $ into individual pump modes the pump
powers of the individual modes are greater for smaller numbers $ K
$ of modes. As a consequence the nonlinear interaction inside a
smaller number of individual triplets is effectively more
developed. This is documented in Fig.~\ref{fig1} for the twin
beams generated by pump pulses with different spectral widths $
\Delta\lambda_{\rm p} $. The analysis presented in
\cite{PerinaJr2015} shows that, for the considered spectral
widths, the wider the spectral width $ \Delta\lambda_{\rm p} $,
the greater the number $ K $ of modes and so the slower the
increase of the number $ N_{\rm s} $ of emitted photon pairs. The
comparison of curves plotted in Fig.~\ref{fig1} for the 4-mm and
8-mm long crystals reveals that the numbers $ N_{\rm s} $ of
emitted photon pairs obtained in the longer crystal are smaller
than those reached in the shorter crystal. This is given by the
fact that individual mode triplets interact longer in the 8-mm
long crystal and thus a greater number of down-converted modes
faces pump depletion. For comparison, only around 1~\% of the
pump-pulse energy is transferred to the down-converted fields of
the 8-mm long crystal for $ P_{\rm p} \approx 0.2 $~W, whereas
around 8~\% of the pump-pulse energy occurs in the down-converted
fields of the 4-mm long crystal.

The curves in Fig.~\ref{fig2} show that the number $ K $ of modes
considered as a function of power $ P_{\rm p} $ reaches minimum
for a certain threshold value $ P_{\rm p,th} $. The comparison of
curves in Figs.~\ref{fig1} and \ref{fig2} reveals that this
threshold power $ P_{\rm p} $ determines a boundary between the
areas with exponential and linear growths of the twin-beam
intensity. Moreover, as discussed in the next subsection it also
characterizes the power at which maximal coherence of the twin
beam is reached. It is worth noting that the number $ K $ of modes
of the 8-mm long crystal is smaller than that of the 4-mm long
crystal assuming the pump-beam parameters fixed. This originates
in more strict phase-matching conditions of the longer crystal.

\subsection{Spectral coherence}

\begin{figure}         
 (a) \resizebox{.85\hsize}{!}{\includegraphics{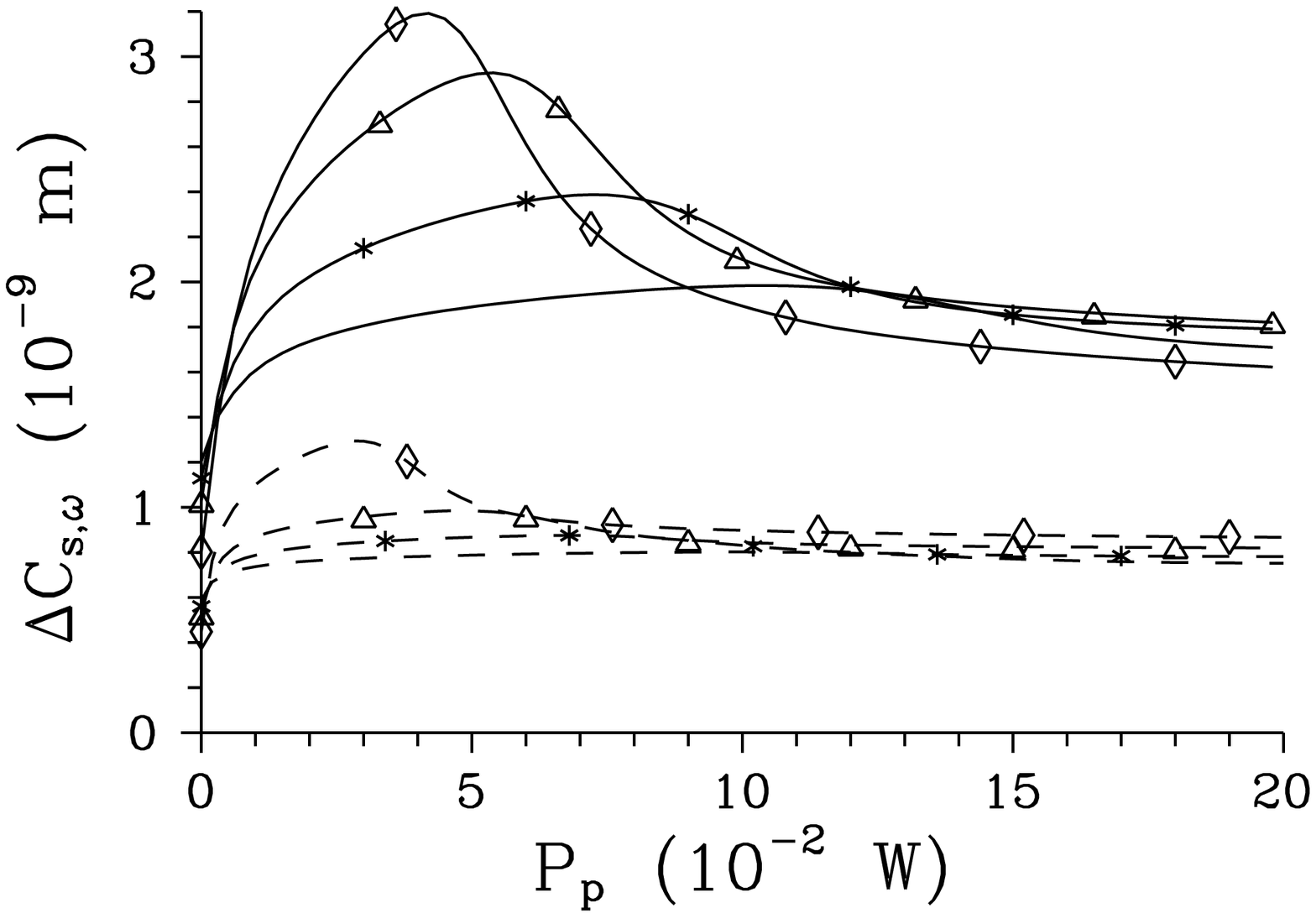}}

  \vspace{2mm}
 (b) \resizebox{.85\hsize}{!}{\includegraphics{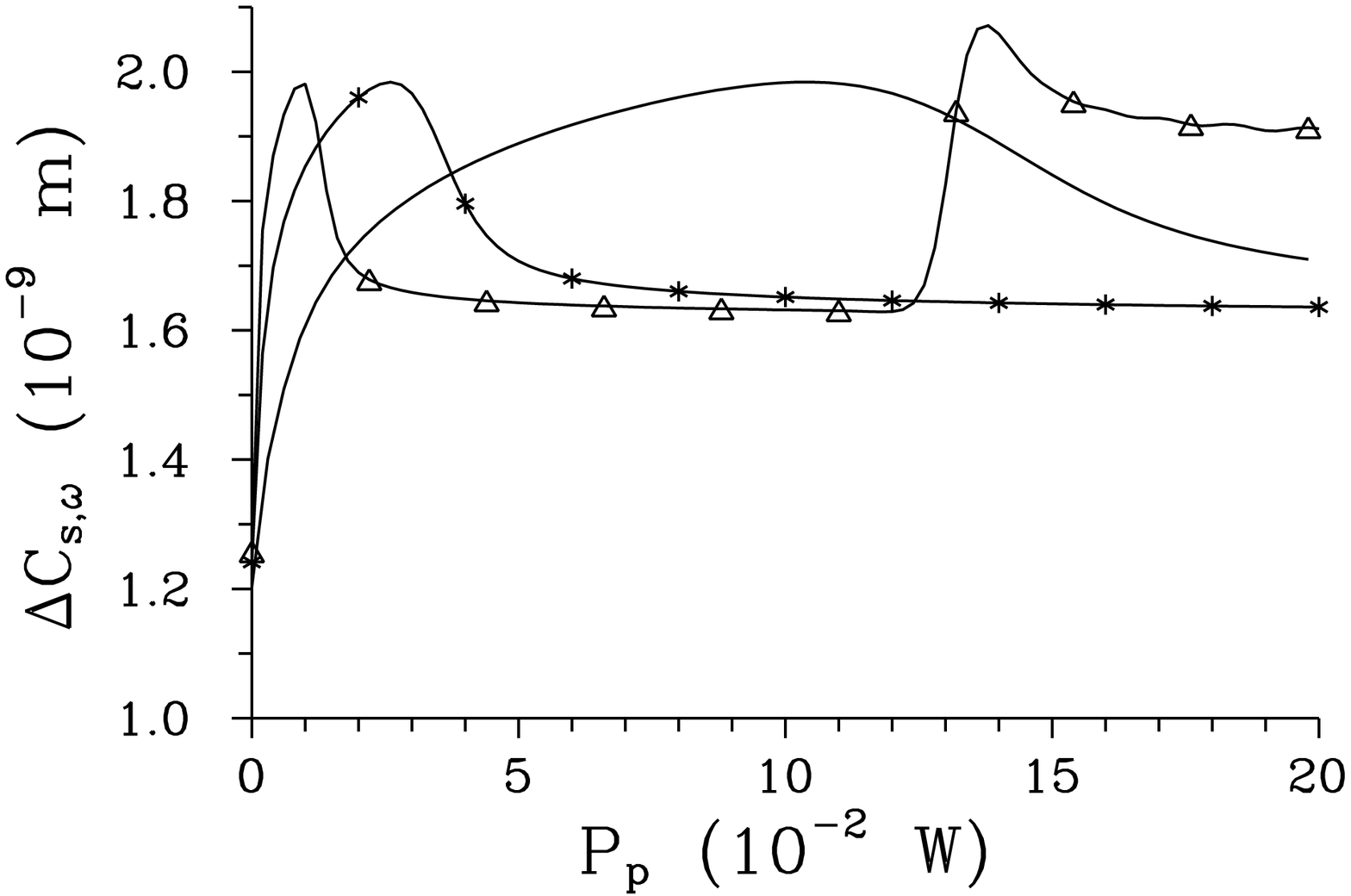}}

  \caption{Widths $ \Delta C_{{\rm s},\omega} $ of spectral intensity cross-correlation functions (FWHM, full width at half maximum)
   as they depend on pump power $ P_{\rm p} $ for 4-mm (solid curves) and 8-mm (dashed curves) long crystals for (a)
   $ \Delta\lambda_{\rm p} = 1 \times 10^{-9} $~m (plain curves),
   $ \Delta\lambda_{\rm p} = 7 \times 10^{-10} $~m (curves with $ \ast $),
   $ \Delta\lambda_{\rm p} = 5 \times 10^{-10} $~m (curves with $ \triangle $),
   and $ \Delta\lambda_{\rm p} = 3 \times 10^{-10} $~m (curves with $ \diamond $);
   $ w_{\rm p} = 1 \times 10^{-3} $~m and (b) $ w_{\rm p} = 1 \times 10^{-3} $~m
   (plain curve), $ w_{\rm p} = 5 \times 10^{-4} $~m (curve with $
   \ast $) and $ w_{\rm p} = 3 \times 10^{-4} $~m (curve with $
   \triangle $); $ \Delta\lambda_{\rm p} = 1 \times 10^{-9} $~m.}
\label{fig3}
\end{figure}

\begin{figure}         
 \resizebox{.85\hsize}{!}{\includegraphics{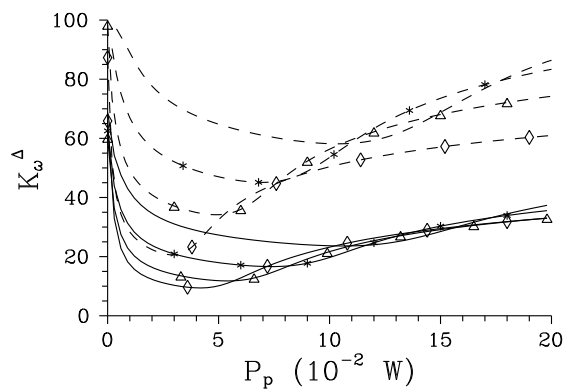}}
  \caption{Numbers $ K_{\omega}^\Delta $ of spectral modes given by the Fedorov ratio
   versus the pump power $ P_{\rm p} $ for 4-mm (solid curves) and 8-mm (dashed curves) long crystals
   for $ \Delta\lambda_{\rm p} = 1 \times 10^{-9} $~m (plain curves),
   $ \Delta\lambda_{\rm p} = 7 \times 10^{-10} $~m (curves with $ \ast $),
   $ \Delta\lambda_{\rm p} = 5 \times 10^{-10} $~m (curves with $ \triangle $),
   and $ \Delta\lambda_{\rm p} = 3 \times 10^{-10} $~m (curves with $ \diamond $);
   $ w_{\rm p} = 1 \times 10^{-3} $~m.}
\label{fig4}
\end{figure}

The number of modes present in a twin beam determines its
coherence. As follows from the comparison of curves in
Figs.~\ref{fig3}(a) and \ref{fig4} giving the widths $ \Delta
C_{{\rm s},\omega} $ of spectral intensity cross-correlation
function and number $ K_{\omega}^\Delta $ of spectral modes for
different pump spectral widths, the greater the number $
K_{\omega}^\Delta $ of spectral modes, the narrower the width $
\Delta C_{{\rm s},\omega} $ of spectral intensity
cross-correlation function. This behavior stems from the
properties of the spectral Schmidt modes. A $q $-th mode has $ q-1
$ minima in its intensity profile. Also, the larger the mode index
$ q $, the more complex the phase profile of the mode. The
increasing power $ P_{\rm p} $ prefers the modes with small
numbers $ q $ (and great values of the Schmidt coefficients $
\lambda_{mlq} $) which naturally leads to the broadening of the
intensity spectral cross-correlation function. When the threshold
value $ P_{\rm r,th} $ of the power is reached, the number $
K_{\omega}^\Delta $ of spectral modes begins to increase, modes
with greater values of index $ q $ become more important and, as a
consequence, the twin-beam spectral coherence decreases. The pump
spectral width $ \Delta\lambda_{\rm p} $ and crystal length $ L $
are two critical parameters determining the spectral coherence of
twin beam. Decrease of the pump spectral width $
\Delta\lambda_{\rm p} $ (until certain width is reached
\cite{PerinaJr2015a,Fedorov2008}) results in the decrease of the
number $ K_{\omega}^\Delta $ of spectral modes, which leads to the
growth of spectral coherence and, hand in hand, to lower values of
the threshold power $ P_{\rm p,th} $.

The spectra of down-converted fields as well as their intensity
auto- and cross-correlation functions for the 8-mm long crystal
are narrower than those of the 4-mm long crystal due to the more
strict phase-matching conditions along the $ z $ axis [compare the
solid and dashed curves in Fig.~\ref{fig3}(a)]. The number $
K_{\omega}^\Delta $ of spectral modes in the longer crystal is
greater compared to the shorter crystal under the same conditions
(compare the solid and dashed curves in Fig.~\ref{fig4}). However,
the overall numbers of spatio-spectral modes of the longer crystal
are smaller (see Fig.~\ref{fig2}) and so the threshold powers $
P_{\rm p,th} $ determined for the longer crystal are smaller.
Also, the longer interaction length of the 8-mm long crystal and
thus the more intense energy transfer in this crystal contribute
to the smaller threshold powers $ P_{\rm p,th} $.

The pump-beam radius $ w_{\rm p} $ influences the spectral
coherence only indirectly, through the number of modes. The number
of transverse modes decreases with the decreasing pump-beam radius
$ w_{\rm p} $ (for details, see \cite{PerinaJr2015}). Individual
modes' triplets then attain greater pump-field energies and the
effect of pump depletion occurs for smaller powers $ P_{\rm p} $.
As a consequence, the maxima in the spectral coherence are reached
for smaller threshold powers $ P_{\rm p, th} $, as documented in
Fig.~\ref{fig3}(b). The maximal attainable value of the width $
\Delta C_{{\rm s},\omega} $ of spectral intensity
cross-correlation function is practically not affected by the
pump-beam radius $ w_{\rm p} $.

For the analyzed range of pump powers $ P_{\rm p} $, the intensity
auto- $ A_{{\rm s},\omega} $ and cross-correlation $ C_{{\rm
s},\omega} $ functions practically coincide \cite{PerinaJr2015a}.
Spectra of the down-converted beams considered as functions of the
power $ P_{\rm p} $ behave in the opposed way than the correlation
functions, i.e. they are becoming narrower with the increasing
power $ P_{\rm p} $ until the threshold power $ P_{\rm p, th} $ is
reached. For greater powers $ P_{\rm p} $ they broaden. This
behavior originates in the fact that the intensity spectral mode
profiles $ |f_{{\rm s},q}(\omega)|^2 $ broaden with the increasing
index $ q $. As the modes with small numbers $ q $ dominate in the
area around the threshold power $ P_{\rm p, th} $, the spectra are
narrower there.

\begin{figure}         
 \resizebox{.85\hsize}{!}{\includegraphics{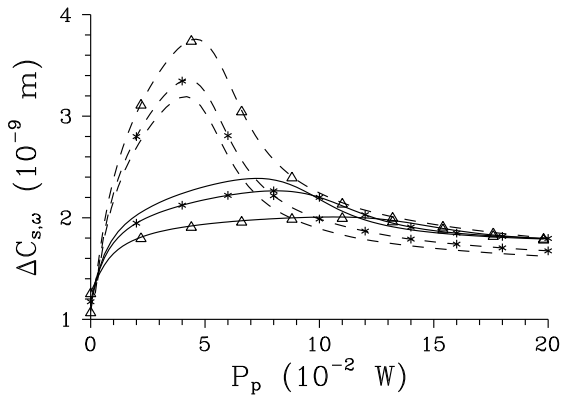}}
  \caption{Widths $ \Delta C_{{\rm s},\omega} $ of intensity cross-correlation functions (FWHM)
   as they depend on pump power $ P_{\rm p} $ for $ \Delta\lambda_{\rm p} = 3 \times 10^{-10} $~m
   (solid curves) and $ \Delta\lambda_{\rm p} = 7 \times 10^{-10} $~m
   (dashed curves) assuming $ a_{\rm p} = 0 $ (plain curves), $ a_{\rm p} = 0.5
   $ (curves with $ \ast $) and $ a_{\rm p} = 1 $ (curves with $ \triangle
   $); $ w_{\rm p} = 1 \times 10^{-3} $~m, $ L = 4 \times 10^{-3}
   $~m.}
\label{fig5}
\end{figure}

Spectral coherence of the twin beam is also affected by the
pump-pulse chirp parameter $ a_{\rm p} $. In general, a nonzero
chirp parameter $ a_{\rm p} $ introduces additional phase
modulation to the two-photon spectral amplitude in the direction
perpendicular to the direction $ \omega_{\rm s} + \omega_{\rm i} =
{\rm const} $. This influences the Schmidt decomposition. If the
pump-field spectrum is sufficiently narrow (and enforces more
strict conditions for the nonlinear interaction compared to the
phase-matching condition along the $ z $ axis), the more complex
phase modulation caused by the chirp results in greater numbers $
K_{\omega}^\Delta $ of spectral modes. That is why the twin-beam
spectral coherence decreases with the increasing chirp parameter $
a_{\rm p} $ and also the threshold power $ P_{\rm p, th} $
increases (see the solid curves in Fig.~\ref{fig5} obtained for
the pump-field spectrum 0.3~nm wide). On the other hand, if the
pump-field spectrum is broader such that the phase-matching
condition along the $ z $ axis causes a larger phase modulation,
interference of this modulation with that coming from the
pump-field chirp occurs. This may result in the increase of
spectral coherence, as documented in Fig.~\ref{fig5} for the
pump-field spectrum 0.7~nm wide.

\subsection{Spatial coherence}

\begin{figure}         
 (a) \resizebox{.85\hsize}{!}{\includegraphics{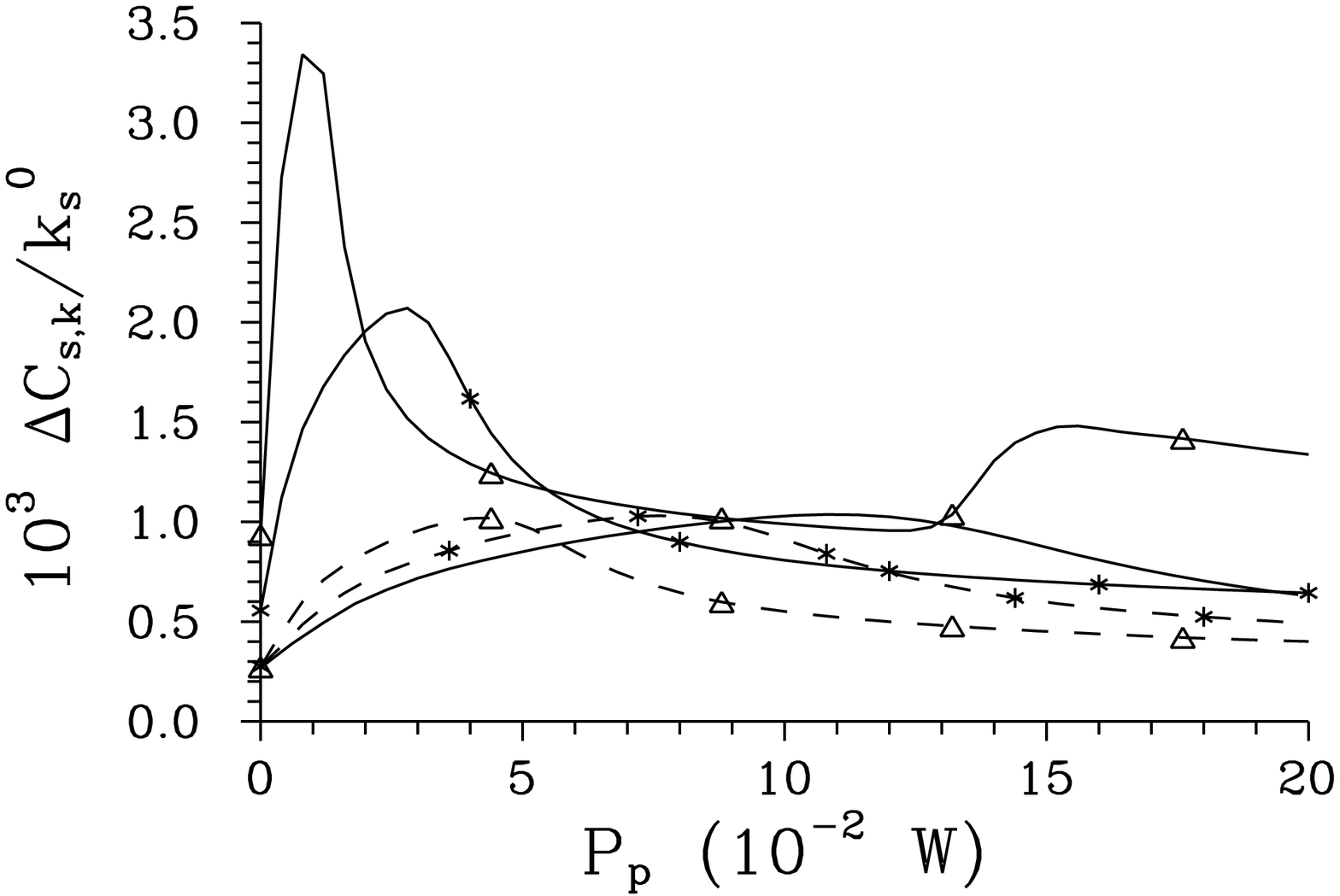}}

  \vspace{2mm}
 (b) \resizebox{.85\hsize}{!}{\includegraphics{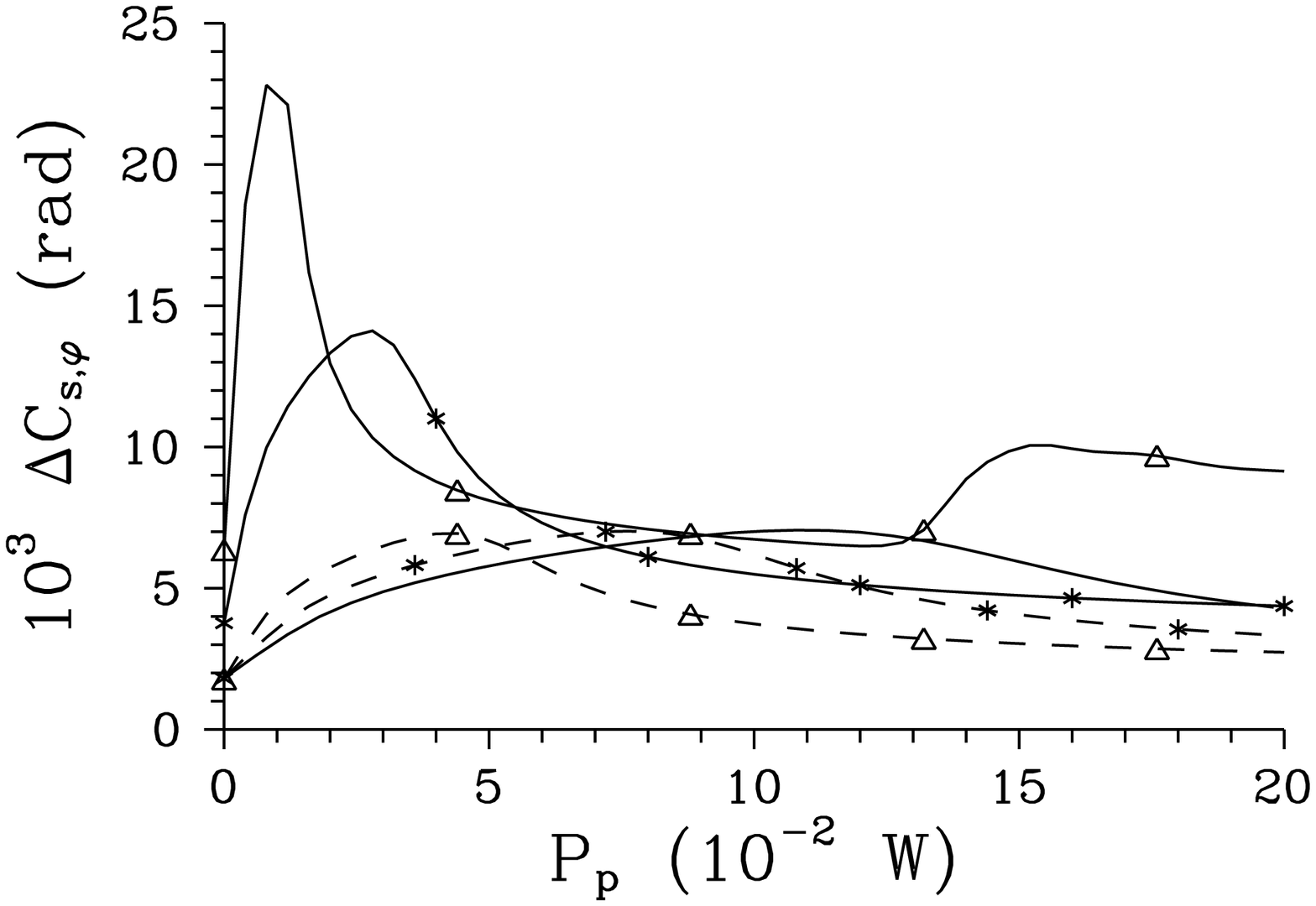}}

  \caption{(a) [(b)] Widths $ \Delta C_{{\rm s},k} $ [$ \Delta C_{{\rm s},\varphi} $] of radial [azimuthal] transverse
   intensity cross-correlation functions (FWHM) as they depend on pump power $ P_{\rm p} $ for
   $ \Delta\lambda_{\rm p} = 1 \times 10^{-9} $~m, $ w_{\rm p} = 1 \times 10^{-3} $~m
   (plain solid curve), $ w_{\rm p} = 5 \times 10^{-4} $~m (solid curve with $
   \ast $), $ w_{\rm p} = 3 \times 10^{-4} $~m (solid curve with $
   \triangle $) and $ w_{\rm p} = 1 \times 10^{-3} $~m,
   $ \Delta\lambda_{\rm p} = 7 \times 10^{-10} $~m (dashed curve with $ \ast $),
   $ \Delta\lambda_{\rm p} = 3 \times 10^{-10} $~m (dashed curve with $ \triangle
   $); $ L = 4 \times 10^{-3} $~m.}
\label{fig6}
\end{figure}

Spatial coherence of the twin beam defined in the wave-vector
transverse plane (far field) behaves qualitatively in the same way
as the spectral coherence provided that we consider the pump-beam
radius $ w_{\rm p} $ instead of the pump-field spectral width $
\Delta\lambda_{\rm p} $. Coherence in the radial as well as
azimuthal directions of the transverse plane increases with the
decreasing pump-beam radius $ w_{\rm p} $ [see Fig.~\ref{fig6}(a)
for the radial direction and Fig.~\ref{fig6}(b) for the azimuthal
direction]. This behavior stems from the decrease of the number of
Schmidt modes in the transverse plane observed for the decreasing
pump-beam radius $ w_{\rm p} $ (for details, see
\cite{PerinaJr2015a}). Similarly as in the spectral domain, this
is accompanied by decreasing values of the threshold power $
P_{\rm p, th} $. Contrary to the maximal widths $ \Delta C_{{\rm
s},k} $ and $ \Delta C_{{\rm s},\varphi} $ of the radial and
azimuthal intensity cross-correlation functions, respectively, the
threshold powers $ P_{\rm p, th} $ depend also on the pump-field
spectral width $ \Delta\lambda_{\rm p} $. As the number $ K $ of
modes decreases with the decreasing spectral width $
\Delta\lambda_{\rm p} $, the threshold power $ P_{\rm p, th} $
also decreases.

It is worth noting that, in Fig.~\ref{fig6}, the azimuthal widths
$ \Delta C_{{\rm s},\varphi} $ are approx. 7-times wider than the
radial widths $ \Delta C_{{\rm s},k} $. This is caused by the
strong influence of the phase-matching condition along the $ z $
axis when determining the radial correlation functions. Despite
this, the influence of pump power $ P_{\rm p} $ to the coherence
is quite similar in both cases. We note that the intensity auto-
and cross-correlation functions nearly coincide in both radial and
azimuthal directions for the analyzed range of powers $ P_{\rm p}
$.

The number of modes in the transverse plane is typically by
several orders in magnitude larger than the number $
K_{\omega}^\Delta $ of spectral modes. On the other hand, the
number of modes in the radial direction is usually of the same
order as the number of spectral modes. Moreover, their profiles
behave in the similar way as the profiles of the spectral modes.
That is why, we observe narrowing of the emission ring with the
increasing power $ P_{\rm p} $ until the threshold value $ P_{\rm
p, th} $ is reached. The ring then widens with the increasing
power $ P_{\rm p} $. Comparing the 4-mm and 8-mm long crystals,
the number of radial modes is larger in the shorter crystal
whereas the number of azimuthal modes is similar in both crystals.

We note that the Fourier transform of mode profiles in the
transverse wave-vector plane gives the Schmidt mode profiles in
the crystal output plane. These modes, however, have a specific
structure analyzed in \cite{PerinaJr2015a} that assures practical
independence of the intensity correlation functions on the pump
power $ P_{\rm p} $.

\subsection{Temporal coherence}

\begin{figure}         
 \resizebox{.85\hsize}{!}{\includegraphics{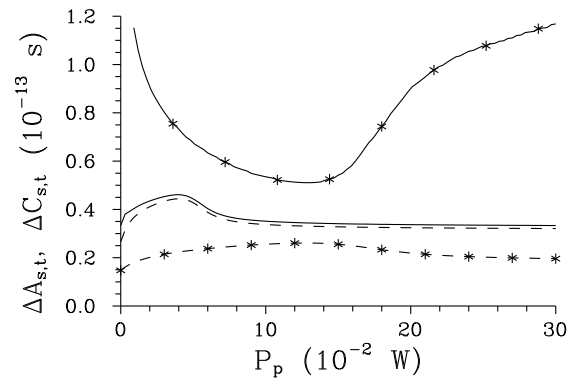}}
  \caption{Widths $ \Delta C_{{\rm s},t} $ (solid curves) and $ \Delta A_{{\rm s},t} $ (dashed curves) of temporal intensity auto- and
   cross-correlation functions (FWHM) as they depend on pump power $ P_{\rm p} $ for
   $ \Delta\lambda_{\rm p} = 3 \times 10^{-10} $~m (plain curves) and
   $ \Delta\lambda_{\rm p} = 1.2 \times 10^{-9} $~m (curves with $ \ast $); $ w_{\rm p} = 1
   \times 10^{-3} $~m, $ L = 4 \times 10^{-3} $~m.}
\label{fig7}
\end{figure}

Temporal intensity auto- and cross-correlation functions mutually
differ. The intensity auto-correlation functions $ A_{{\rm s},t} $
are in general narrower than their cross-correlation counterparts
$ C_{{\rm s},t} $ (see Fig.~\ref{fig7}). Their behavior with
respect to the change of pump power $ P_{\rm p} $ depends on the
pump-field spectral width $ \Delta\lambda_{\rm p} $. Provided that
the pump-field spectral width $ \Delta\lambda_{\rm p} $ is
sufficiently small, the auto- and cross-correlation functions
behave in the same way as their spectral counterparts. Both of
them exhibit their maxima at the threshold pump power $ P_{\rm p,
th} $ determined in the spectral domain [compare the curves in
Figs.~\ref{fig7} and \ref{fig3}(a) for $ \Delta\lambda_{\rm p} = 3
\times 10^{-10} $~m]. This behavior can be explained in the same
way as in the spectrum because the profiles of temporal modes are
similar to those of the spectral modes. We note that the
Schmidt-mode structure and the varying mean photon numbers of the
modes suppress the natural characteristic of the Fourier transform
that assigns longer pulses to narrower spectra.

However, when the pump-field spectral width $ \Delta\lambda_{\rm
p} $ is wider so that the nonlinear phase mismatch introduces
greater phase modulations in the two-photon spectral amplitude, we
observe decrease in the width $ \Delta C_{{\rm s},t} $ of
intensity cross-correlation function for the pump powers around $
P_{\rm p, th} $ (see the solid curve drawn for $
\Delta\lambda_{\rm p} = 1.2 \times 10^{-9} $~m in
Fig.~\ref{fig7}). In this area, the twin beam is composed of a
reduced number of spectral modes that have the greatest Schmidt
coefficients $ \lambda $ and exhibit only mild phase modulation.
The strong phase spectral modulation described in a weak twin beam
by the modes with smaller Schmidt coefficients are suppressed thus
allowing shortening of the temporal cross-correlations. The
intensity auto-correlation function $ A_{{\rm s},t} $ is not
sensitive to the phase modulation and so it evolves with the pump
power $ P_{\rm p} $ in the same manner as its spectral
counterpart. Also, the signal-field pulse duration shortens with
the increasing power $ P_{\rm p} $ up to its threshold value $
P_{\rm p, th} $, where it begins to lengthen.

\section{Multiple coherence maxima}

\begin{figure}         
 \resizebox{.85\hsize}{!}{\includegraphics{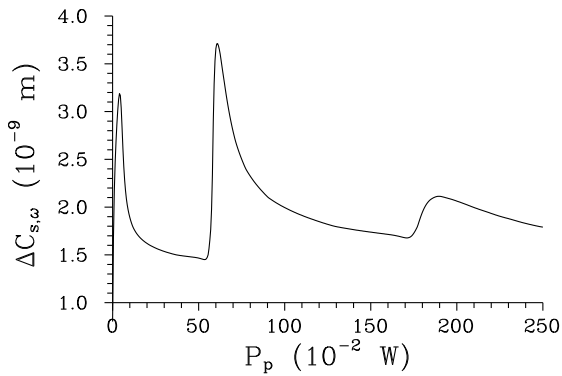}}
  \caption{Width $ \Delta C_{{\rm s},\omega} $ of spectral intensity cross-correlation functions (FWHM)
   as it depends on pump power $ P_{\rm p} $; $ \Delta\lambda_{\rm p} = 3 \times 10^{-10} $~m,
   $ w_{\rm p} = 1 \times 10^{-3} $~m, $ L = 4 \times 10^{-3} $~m.}
\label{fig8}
\end{figure}

\begin{figure}   
 \resizebox{.61\hsize}{!}{\includegraphics{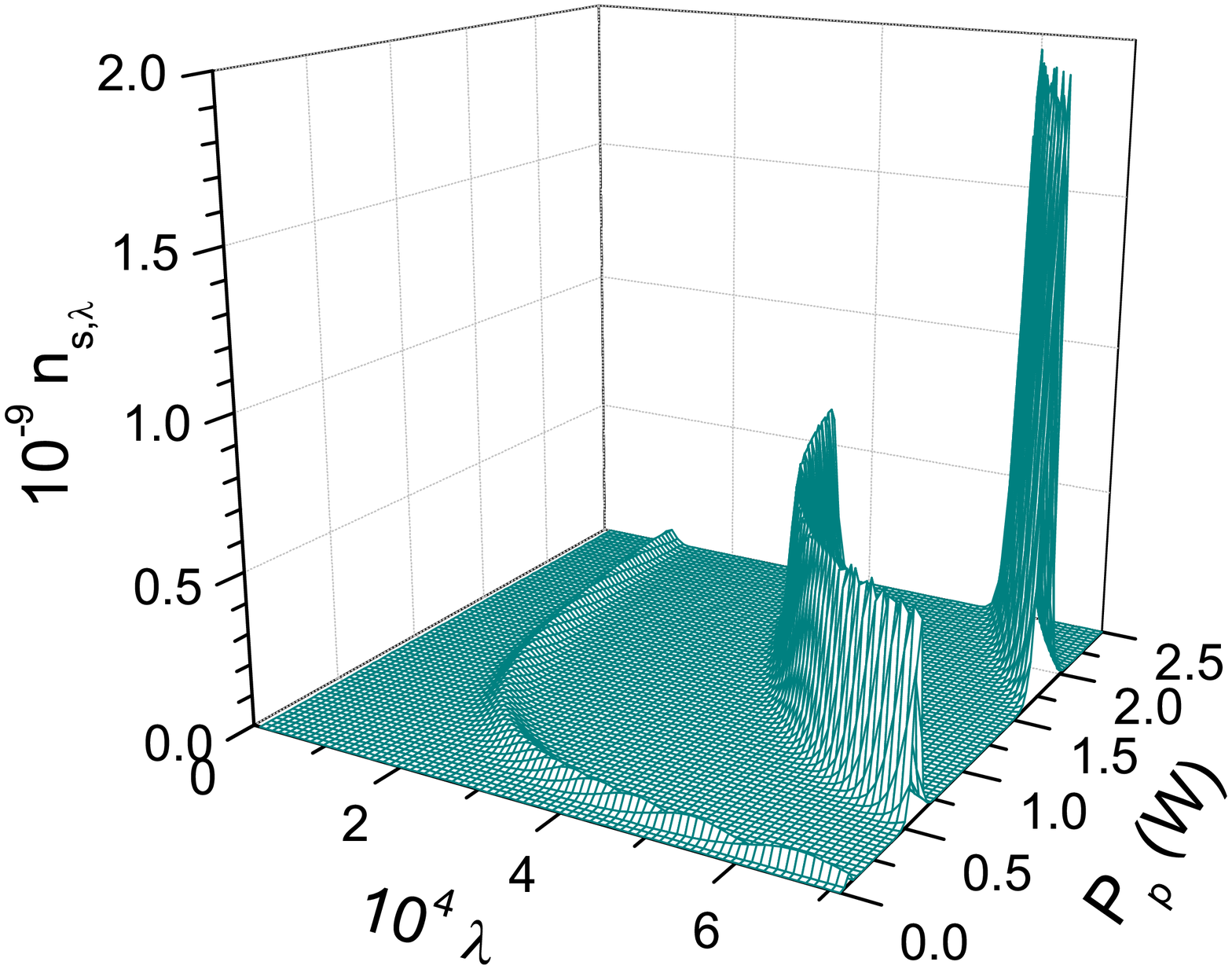}}
 \resizebox{.32\hsize}{.48\hsize}{\includegraphics{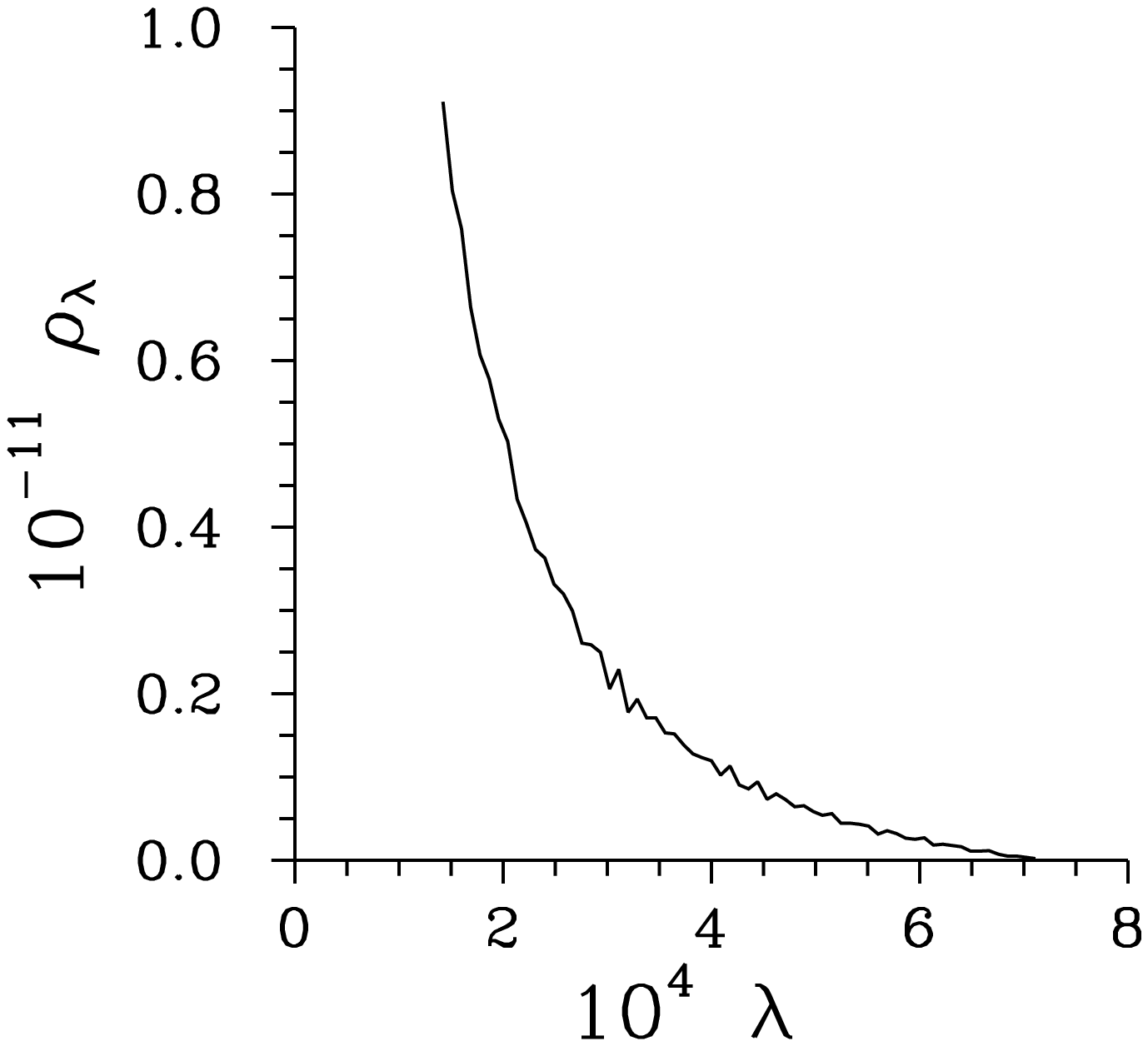}}

 \centering{(a) \hspace{0.35\hsize} (b) }

 \caption{(Color online) (a) Mean number $ n_{{\rm s},\lambda} $ of signal photons in modes with the Schmidt
 coefficients $ \lambda $ as they depend on pump power $ P_{\rm p} $ and (b) density $ \varrho_\lambda $ of
 modes revealed by the Schmidt decomposition; $ n_{\lambda} = \langle \hat{a}_{{\rm s},mlq}^\dagger
 \hat{a}_{{\rm s},mlq} \rangle $ for $ mlq $ such that $ \lambda_{ml}\lambda_q = \lambda $;
 parameters are given in the caption to Fig.~\ref{fig8}.}
\label{fig9}
\end{figure}

\begin{figure}         
 \resizebox{.85\hsize}{!}{\includegraphics{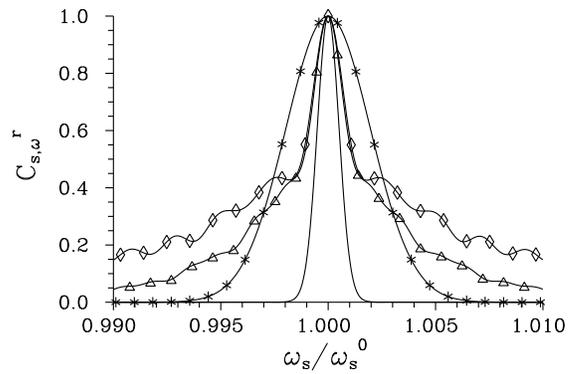}}
  \caption{Spectral intensity cross-correlation functions $ C_{{\rm s},\omega}^{\rm r}(\omega_{\rm s}) \equiv
  C_{{\rm s},\omega}(\omega_{\rm s},\omega_{\rm i}^0)/
  C_{{\rm s},\omega}(\omega_{\rm s}^0,\omega_{\rm i}^0) $ are plotted for
  $ P_{\rm p} = 1 \times 10^{-7} $~W (plain curve), $ P_{\rm p} = 4 \times 10^{-2} $~W
  (curve with $ \ast $); $ P_{\rm p} = 6.1 \times 10^{-1} $~W
  (curve with $ \triangle $), and $ P_{\rm p} = 1.3 $~W
  (curve with $ \diamond $);  parameters are given in the caption to Fig.~\ref{fig8}.}
\label{fig10}
\end{figure}

The origin of coherence maxima discussed above and observed
simultaneously in the spectrum (see Fig.~\ref{fig8}) and radial
and azimuthal directions in the wave-vector transverse plane has
been explained by the reduction of the number of spatio-spectral
twin-beam modes.
This reduction is a consequence of different evolution of the mean
photon numbers in the down-converted modes belonging to different
modes' triplets. The modes with the greatest Schmidt coefficients
$ \lambda $ take energy from the corresponding pump modes faster
and so they grow more rapidly. That is why, they completely
deplete their pump modes for smaller powers $ P_{\rm p} $ (keeping
the crystal length fixed) and reach their maximal mean photon
numbers $ n_{{\rm s},\lambda} $ [see Fig.~\ref{fig9}(a)].
For larger powers $ P_{\rm p} $, the down-converted modes reach
their maximal photon numbers inside the crystal and they return
certain part of their energy back to their pump modes before
leaving the crystal. For certain pump power depending on the
triplet, the down-converted modes of the appropriate triplet leave
the crystal in the vacuum state. Above this power, the triplet's
evolution repeats from the beginning. As a consequence, the modes
with the greatest Schmidt coefficients $ \lambda $ again reach
their maximal mean photon numbers $ n_{{\rm s},\lambda} $ for a
sufficiently high power $ P_{\rm p, th1} $. These maximal photon
numbers are now considerably greater than those reached for $
P_{\rm p, th} $ as the incident power $ P_{\rm p} $ is greater and
the mode structure is assumed independent on the pump power. Owing
to the same reason as discussed for the powers around $ P_{\rm p,
th} $, there again occur local maxima in the spectral and spatial
intensity correlation functions. As the incident pump powers
belonging to individual modes' triplets are greater compared to
those found around the power $ P_{\rm p, th} $, the effect of
modes' number reduction is even stronger and so even wider
intensity correlation functions are observed. However, the
coherence properties in this case are influenced also by another
group of modes with smaller Schmidt coefficients [see
Fig.~\ref{fig9}(a)]. These modes have much lower mean photon
numbers compared to the modes with the greatest Schmidt
coefficients on one hand, on the other hand their number is much
larger [see the density $ \varrho_\lambda $ of modes plotted in
Fig.~\ref{fig9}(b)]. They form a narrower peak in the intensity
cross-correlation function $ C_{{\rm s},\omega} $, on the top of a
broader peak created by the modes with the greatest Schmidt
coefficients $ \lambda $ (see Fig.~\ref{fig10}). For the range of
powers $ P_{\rm p} $ investigated in Fig.~\ref{fig8}, we even
observe the third threshold power $ P_{\rm p, th2} $. As three
different groups of modes coexist in the twin beam for the powers
around $ P_{\rm p, th2} $, the coherence peak is less pronounced.
Also, the profile of intensity cross-correlation function $
C_{{\rm s},\omega} $ is more complex. This behavior is illustrated
in Fig.~\ref{fig10} where the profile of cross-correlation
function $ C_{{\rm s},\omega} $ is drawn for the pump powers $
P_{\rm p} \approx 0 $~W, $ P_{\rm p, th} $, $ P_{\rm p, th1} $,
and $ P_{\rm p, th2} $.

\section{Model with the extended crystal length}

In the crystals like BBO in the considered type-I interaction
configuration there occurs walk-off of the interacting fields due
to anisotropy of the pump beam \cite{Fedorov2008}. Owing to the
walk-off, the modes' triplets interacting at the beginning of the
crystal gradually loose their mutual overlap in the transverse
plane which effectively weakens the nonlinear interaction. On the
other hand, new modes' triplets can be seeded as the pump beam
propagates along the crystal. This poses the question about an
effective interaction length for individual modes' triplets and
also about an appropriate crystal length that determines the
Schmidt-mode profiles. We note that the shorter the crystal, the
wider the spectral and radial spatial modes. This affects
coherence of the twin beam. The developed model is not able to
incorporate the influence of crystal anisotropy directly. On the
other hand, we may consider two different crystal lengths in the
model. The first length $ L $ determines the extension of mode
profiles whereas the second length $ L_{\rm ext} $ characterizes
an effective length of the nonlinear interaction.

\begin{figure}         
 \resizebox{.85\hsize}{!}{\includegraphics{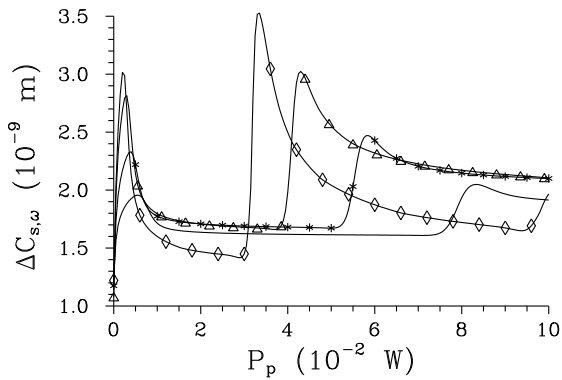}}
  \caption{Widths $ \Delta C_{{\rm s},\omega} $ of spectral intensity cross-correlation functions
  (FWHM) as they depend on pump power $ P_{\rm p} $ for $ \Delta\lambda_{\rm p} = 1 \times 10^{-9} $~m (plain curve),
   $ \Delta\lambda_{\rm p} = 7 \times 10^{-10} $~m (curves with $ \ast $),
   $ \Delta\lambda_{\rm p} = 5 \times 10^{-10} $~m (curves with $ \triangle $),
   and $ \Delta\lambda_{\rm p} = 3 \times 10^{-10} $~m (curves with $ \diamond $);
   $ w_{\rm p} = 5 \times 10^{-4} $~m, $ L = 4 \times
   10^{-3} $~m, $ L_{\rm ext} = 8 \times 10^{-3} $~m.}
\label{fig11}
\end{figure}

To reveal the main features of the twin beams in such a model, we
consider a crystal 8-mm long and twin-beam modes appropriate for a
4-mm long crystal. Widths $ \Delta C_{{\rm s},\omega} $ of
spectral intensity cross-correlation functions depending on the
pump power $ P_{\rm p} $ are plotted in Fig.~\ref{fig11} for
several pump-field spectral widths $ \Delta\lambda_{\rm p} $. The
comparison of widths $ \Delta C_{{\rm s},\omega} $ drawn in
Fig.~\ref{fig11} with those plotted in Fig.~\ref{fig3}(a) for the
crystal 4-mm long reveals that the maximal spectral widths $
\Delta C_{{\rm s},\omega} $ are approximately the same for both
cases. However, the threshold pump powers $ P_{\rm p,th} $ are
more than 4-times lower for the 8-mm long crystal. This arises
from the fact that the interaction strength scales as $
L\sqrt{P_{\rm p}} $. In other words, a crystal two times longer
needs only one quarter of the incident power to provide the same
interaction. The vacuum contributions to the evolution of modes'
triplets even move the threshold powers $ P_{\rm p, th} $ to
slightly lower values. For this reason, we observe the second
threshold powers $ P_{\rm p, th1} $ in Fig.~\ref{fig11} for the
pump powers $ P_{\rm p} $ considerably lower than in the case of
the 4-mm long crystal.

\section{Conclusions}

Using the spatio-spectral Schmidt dual signal and idler modes and
the corresponding pump modes, we have developed the model of
intense parametric down-conversion applicable in the regime with
pump depletion. Contrary to the usual parametric approximation,
the model assumes a classical pump beam that undergoes depletion
during its propagation (\emph{the generalized parametric
approximation}). Due to the pump depletion, the initial increase
of spectral, temporal and transverse wave-vector coherence on the
pump-power axis is replaced by a decrease. This occurs as a
consequence of the back-flow of energy from the down-converted
modes into the pump modes observed in the most strongly
interacting modes' triplets (with the greatest Schmidt
coefficients). The change of twin-beam coherence reflects the
relative change of mean photon numbers in individual twin-beam
modes. The better the coherence the smaller the number of
well-populated modes and vice versa. The threshold pump power at
which the best coherence is reached has been analyzed as it
depends on the pump spectral width and spatial radius. The
relationship between the coherence and the twin-beam mode
structure predicts the existence of additional threshold pump
powers at which the local coherence functions reach their maxima.
The curves obtained for the 4-mm and 8-mm long BBO crystals in the
typical experimental configuration have provided a deeper insight
into the twin-beam properties useful, e.g., in interpreting the
experimental results.

\acknowledgments

The author thanks M. Bondani, J. Pe\v{r}ina, O. Haderka, A. Allevi
and O. Jedrkiewicz for stimulating discussions. He gratefully
acknowledges the support by project 15-08971S of the Grant Agency
of the Czech Republic and project LO1305 of the Ministry of
Education, Youth and Sports of the Czech Republic.

\bibliography{perina}
\bibliographystyle{unsrt}  

\end{document}